\documentclass[%
 aip,
 amsmath,amssymb,
 reprint,%
]{revtex4-1}
\newcommand{\bvec}[1]{\mathbf{#1}}
\newcommand{\eq}[2]{\begin{equation}#1\label{#2}\end{equation}}

\newcommand{\moire}{ moir\'{e} }

\usepackage{graphicx}% Include figure files
\usepackage{dcolumn}% Align table columns on decimal point
\usepackage{bm}% bold math
\usepackage[mathlines]{lineno}% Enable numbering of text and display math
\usepackage[utf8]{inputenc}
\usepackage[T1]{fontenc}
\usepackage{mathptmx}
\usepackage{etoolbox}
\usepackage{ulem}
\usepackage{wasysym}
\usepackage[english]{babel}

\makeatletter
\def\@email#1#2{%
 \endgroup
 \patchcmd{\titleblock@produce}
  {\frontmatter@RRAPformat}
  {\frontmatter@RRAPformat{\produce@RRAP{*#1\href{mailto:#2}{#2}}}\frontmatter@RRAPformat}
  {}{}
}%
\makeatother

\begin{document}

% \preprint{AIP/123-QED}

\title{PyAtoms: An interactive tool for simulating atomic scanning tunneling microscopy images of 2D materials, moiré systems and superlattices}
% Force line breaks with \\

\author{A.G. Prado}
\author{M.I. Mandigo-Stoba}
\author{K.-Y. Wey}
\author{S. Nekarae}
\author{A. Enriquez-Ibarra}
\author{S. Ba\~{n}uelos}
\author{A. Nguyen}
\author{C. Guti\'{e}rrez}
\email{gutierrez@physics.ucla.edu.}

\affiliation{ 
Department of Physics and Astronomy, University of California, Los Angeles, Los Angeles, California, 90095 USA%\\This line break forced with \textbackslash\textbackslash
}%

% \author{C. Author}
%  \homepage{http://www.Second.institution.edu/~Charlie.Author.}
% \affiliation{%
% Second institution and/or address%\\This line break forced% with \\
% }%


\begin{abstract}
We present PyAtoms, an interactive open-source software that rapidly simulates atomic-scale scanning tunneling microscopy (STM) and other scanning probe microscopy (SPM) images of two-dimensional (2D) layered materials, moir\'{e} systems, and superlattices. Rooted in a Fourier-space description of ideal atomic lattice images, PyAtoms is a Python-based graphical user interface (GUI) with robust capabilities for tuning lattice parameters (lattice constants, strain, number of layers, twist angles) and STM imaging parameters (pixels, scan size, scan angle) and provides time estimates for spectroscopic measurements. These capabilities allow users to efficiently plan time-consuming STM experiments. We provide an overview of PyAtoms' current features, describe its underlying mathematical principles, and then demonstrate simulations of several 2D materials including graphene with variable sublattice asymmetry, twisted tri-layer graphene moir\'{e} systems, and several charge- and bond-density wave systems.
\end{abstract}

\maketitle

% \begin{quotation}
% The ``lead paragraph'' is encapsulated with the \LaTeX\ 
% \verb+quotation+ environment and is formatted as a single paragraph before the first section heading. 
% (The \verb+quotation+ environment reverts to its usual meaning after the first sectioning command.) 
% Note that numbered references are allowed in the lead paragraph.
% %
% The lead paragraph will only be found in an article being prepared for the journal \textit{Chaos}.
% \end{quotation}

\section{Introduction}
Scanning tunneling microscopy (STM) \cite{binnig_scanning_1987} is a powerful technique for investigating the atomic-scale electronic and structural properties of materials \cite{moler_imaging_2017}. In STM, an atomically sharp tip is rastered across a conductive sample surface while recording the current of electrons tunneling from the tip to the sample, or vice versa. The tunneling current is proportional to the energy-integrated local density of states (LDOS), and can thus provide valuable insights into the spatially-dependent electronic properties of novel quantum materials such as high-temperature superconductors \cite{Fischer_RevModPhys.79.353,hoffman_imaging_2002}, charge- or bond-density wave systems \cite{Sacks_PhysRevB.57.13118,soumyanarayanan2013quantum,arguello_visualizing_2014,gutierrez_imaging_2016,gye2019topological,pasztor_holographic_2019}, and two-dimensional (2D) van der Waals structures and moir\'{e} superlattices \cite{rutter2007scattering,xue2011scanning,decker_local_2011,zhang_interlayer_2017,kerelsky_maximized_2019,jiang_charge_2019,xie_spectroscopic_2019,li_imaging_2021,kim_imaging_2023,nuckolls_quantum_2023,slot2023quantum,ganguli2023visualization}.

A drawback of STM measurements is that they are notoriously time-consuming: Depending on the particular imaging mode (topography or spectroscopy), the time to complete a measurement can take minutes to several \textit{days}. In a STM topography measurement, the STM tip rapidly scans the surface (typical scan rate of $\sim 1$ Hz) at a single tunneling energy while recording the tunnel current line-by-line; in a STM spectroscopy measurement, the tunnel current and/or the differential conductance, $dI_t/dV_b\propto$ LDOS($eV_b$), (where $I_t$ is the tunnel current, $V_b$ is the voltage bias, and $e$ is the elementary charge), is \textit{slowly} measured pixel-by-pixel at one or several tunneling energies. The choice of spatial resolution is especially important for spectroscopic measurements, where the Fourier-transformed STM (FT-STM) image can display intricate electronic scattering patterns -- called quasi-particle interference (QPI) patterns -- that are directly related to constant energy cuts of the momentum-resolved electronic band structure \cite{hoffman_imaging_2002,avraham_quasiparticle_2018}. The significant time commitment of STM imaging in real- or reciprocal-space thus makes it absolutely crucial for researchers to determine the suitable imaging parameters \textit{prior} to engaging a long-term measurement. Recently, other scanning probe microscopy (SPM) techniques such as piezoresponse force microscopy (PFM) and torsional force microscopy (TFM) performed in ambient conditions have become widely used tools for rapid characterization and quality control of 2D moiré samples\cite{mcgilly_visualization_2020,pendharkar_torsional_2024}. These techniques share the need for careful selection of the real-space parameters, e.g. scan range, required to capture the essential features for analyzing the data in reciprocal-space.  Additionally, multilayer moiré samples have the added difficulty of interpreting images which may have several or many intersecting length scales in the form of atomic moiré and supermoiré patterns. Efficient planning of STM/SPM measurements would require a tool that could easily simulate STM data in real- and Fourier-space for a variety of quantum material families while also providing real-time estimation of the measurement time.

To address this need, we created PyAtoms: an open-source, Python-based graphical user interface (GUI) equipped with robust capabilities for quickly and easily simulating STM images of several types of 2D van der Waals quantum materials, including graphene, transition metal dichalcogenides, and systems with superlattices such as charge density waves and multi-layered moir\'{e} lattices. PyAtoms relies on a simplified model of an ideal atomic lattice, enabling the simulation of STM images for several quantum materials and allowing users to tune lattice strain and sublattice asymmetries. These capabilities allow PyAtoms to rapidly simulate even complex STM and FT-STM moir\'{e} patterns in real-time in order to, for instance, optimize scanning parameters before a time-consuming measurement or quickly determine local twist and/or lattice mismatch by side-by-side comparison with experimental data. PyAtoms is thus a versatile and accessible software tool for scanning probe and 2D material researchers.

To our knowledge, no other software exists that combines PyAtoms' level of 2D quantum material STM image simulation with a real-time GUI. Gwyddion \cite{Necas2012}, a robust scanning probe microscopy (SPM) data analysis software, contains a module for simulating well-known surfaces (square, triangular, silicon $7\times 7$), but it lacks real-time simulation of multi-layer systems or for tuning strain or sublattice strength. Stephens and Hollen have shared a Matlab GUI for simulating\moire patterns using dot lattices that simultaneously display the Fourier transform \cite{stephens_moire_nodate}. de Jong has shared a very impressive lattice and\moire atomic simulator that allows for tuning the symmetry, strain, and sublattice, but it lacks a real-time GUI \cite{de_jong_moire_2021}. Finally, while PyAtoms is based on idealized atomic lattices, the National Institute of Standards and Technology (NIST) Joint Automated Repository for Various Integrated Simulations (JARVIS) database hosts a robust library of STM images calculated from density functional theory (\url{https://jarvis.nist.gov/jarvisstm}) \cite{choudhary_computational_2021}. 

PyAtoms is built on Python 3 and utilizes widely used libraries such as NumPy \cite{harris_array_2020}, SciPy \cite{virtanen_scipy_2020}, Matplotlib \cite{hunter_matplotlib_2007} and PyQt5 (\url{https://www.riverbankcomputing.com/software/pyqt}). These libraries are readily available through common platforms such as Anaconda (\url{https://www.anaconda.com}). PyAtoms is hosted on Github (\url{https://github.com/asariprado/PyAtoms}) and is regularly updated. PyAtoms runs on Windows, MacOS, and Linux systems.

The article is organized as follows: We first describe the theoretical underpinnings of PyAtoms. We describe how it generates single lattices of various geometries (square, triangular); how we approximate images of\moire and superlattice systems; and how strain and low-pass filtering affect the generated lattice images. We then describe the PyAtoms graphical interface before concluding with its applications in research.

\section{Methods}

 \subsection{The ideal atomic lattice}
The topographic intensity, $T$, of an ideal crystal at position $\bvec{r} = (x, y)$ can be expressed as a Fourier sum \cite{hytch_quantitative_1998,lawler_intra-unit-cell_2010,hamidian_detection_2016},
\eq{T(\bvec{r}) = \sum_{k} A_k e^{i\bvec{g}_k\cdot (\bvec{r}-\bvec{r}_0)},}{main}
\noindent where $\bvec{g}_k$ corresponds to the $k^{th}$ reciprocal lattice vector corresponding to a Bragg reflection, and $A_k$ are the corresponding Fourier amplitudes for the $k^{th}$ mode which, in general, may be complex \cite{hytch_quantitative_1998}, and $\bvec{r}_0$ is the location of the maxima of a reference atom in the image. For the sake of brevity in the following, we assume this reference atom is located at the origin $\bvec{r}_0=(0,0)$.
For a real image ($A_k = A_{-k}$), the image intensity is thus given by:
\eq{T(\bvec{r}) = \sum_{k}2A_k \cos(\bvec{g}_k \cdot \bvec{r}).}{Eqn2} 
For a square lattice with lattice constant $a$, there are two reciprocal lattice vectors:
\eq{\bvec{g}^{sq}_1=\frac{2\pi}{a}\hat{x},  
	\quad \bvec{g}^{sq}_2=\frac{2\pi}{a}\hat{y}}{}\\
For a triangular lattice, the reciprocal lattice vectors are:
\eq{
	\begin{split}
		\bvec{g}^{tri}_1 &= \frac{2\pi}{a}\Big[\hat{x}+\frac{1}{\sqrt{3}}\hat{y}\Big], 
		\quad \bvec{g}_2^{tri}= \frac{2\pi}{a}\Big[-\hat{x}+\frac{1}{\sqrt{3}}\hat{y}\Big], \\
		\quad \bvec{g}^{tri}_3 &= -(\bvec{g}^{tri}_1 + \bvec{g}^{tri}_2).
	\end{split}
}{}

\noindent The ideal square lattice is thus given by
\eq{T_{\square} (\bvec{r}) = N_1 + N_2 [\cos(\bvec{g}_1^{sq} \cdot\bvec{r}) + \cos(\bvec{g}_2^{sq}\cdot\bvec{r})],}{Eq_sq}

\noindent and the ideal triangular lattice can then similarly be given by
\eq{T_{\triangle}(\bvec{r})=N_1 + N_2\sum_{k=1}^3\cos(\bvec{g}_{k}^{tri} \cdot\bvec{r}),}{Eq_tri}
Here, $N_i$ are constants ensuring the normalization of the images, such that $0\leq T (\bvec{r})\leq 1$.
Including higher harmonics of $\bvec{g}_k$ (i.e. more Bragg peaks) in the sum in Eqs. \ref{Eq_sq}, \ref{Eq_tri} will produce atoms that are more localized in space, approaching Dirac delta functions as the number of harmonics increases \cite{stephens_moire_nodate,latychevskaia_moire_2019,de_jong_moire_2021}.

To simulate an arbitrary rotation, $\theta$, of the scan axis or a twist angle between two layers, we use the standard 2D rotation matrix:
\eq{\mathbf{R}_\theta = \begin{pmatrix}
		\cos\theta & -\sin\theta \\
		\sin\theta & \cos\theta
\end{pmatrix}}{RotationMatrix}
To apply a uniform rotation by $\theta$ to the lattice, each of the reciprocal lattice vectors, $\bvec{g}_k$, is rotated to new positions, $\bvec{g}_k'$, individually:
\eq{\bvec{g}_k' = \mathbf{R}_\theta \bvec{g}_k}{}

\begin{figure}
	\centering
	\includegraphics[width=\linewidth]{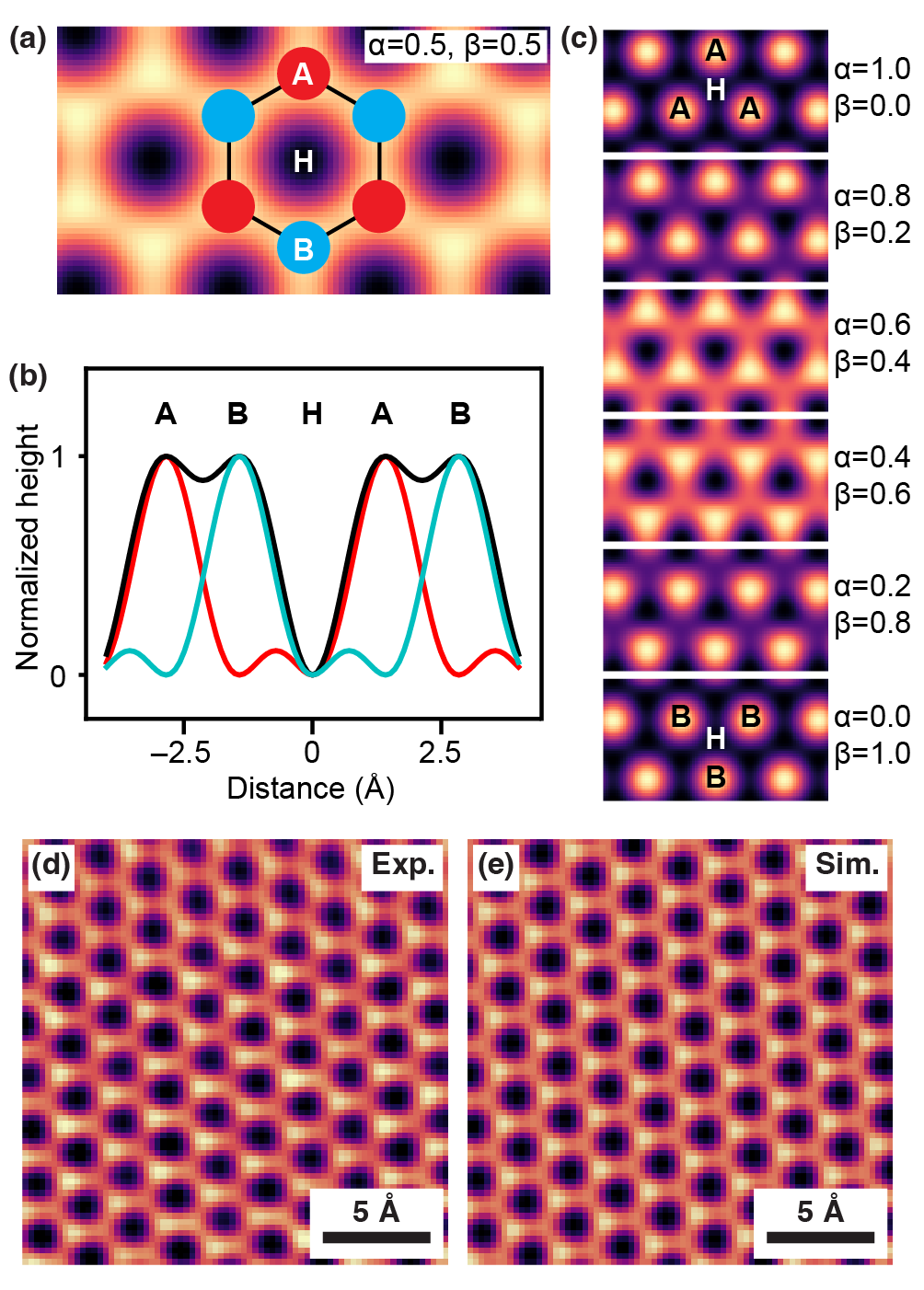}
	\caption{\textbf{Generalized triangular lattices.} (a) PyAtoms simulated image of graphene ($a = 2.46$ \AA) created with equal amplitude sublattices ($\alpha=\beta=0.5$). (b) Vertical line scan of the image in (a) connecting consecutive hollow sites. (c) Series of simulated generalized honeycomb crystals with variable amplitudes on each sublattice. (d) Experimental STM image ($V_b = -400$ mV, $I_t = 250$ pA, $T = $ 4.7 K) of graphite displaying sublattice asymmetry. (e) PyAtoms simulation of (d) ($\alpha = 0.45,\beta = 0.55$) that faithfully reproduces the sublattice asymmetry.} 
	\label{fig:sublattice}
\end{figure}

\subsection{Honeycomb crystal structures}

The honeycomb crystal, such as graphene, consists of a triangular Bravais lattice with a two-atom basis. It can be easily pictured as two interpenetrating triangular Bravais lattices, labeled $A$ and $B$ (Fig. \ref{fig:sublattice}(a)). Ignoring normalization for the moment, the $A$ sublattice centered at the origin, $\bvec{r}_0=(0,0)$, and with amplitude $\alpha$ is given by
\eq{T_A = \alpha \sum_{k} \exp(i\bvec{g}^{tri}_k \cdot\bvec{r}).}{Eq4}
The $B$ sublattice centered at the nearest-neighbor distance $\bvec{s}=(a/\sqrt{3})\hat{y}$ with respect to the $A$ sublattice, and with amplitude $\beta$, is given by
\eq{T_B = \beta \sum_{k} \exp[i\bvec{g}^{tri}_k \cdot(\bvec{r}+\bvec{s})]
	=\beta e^{2\pi i/3}\sum_{k} \exp(i\bvec{g}^{tri}_k \cdot\bvec{r}).}{Eq5}
Here, we use the fact that $\bvec{g}_k^{tri}\cdot\bvec{s}=2\pi/3$ mod $2\pi$ for each $k$. The most general honeycomb crystal (centered at an $A$ sublattice site) is thus given by the sum,
\eq{T_{AB} = T_A + T_B = (\alpha + \beta e^{-2\pi i/3})\sum_{k} \exp(i\bvec{g}_k^{tri} \cdot\bvec{r}).}{Eq6}
The properly normalized atomic honeycomb image is then expressed as:
\eq{T_{\varhexagon} = N_1 + N_2\Re\{T_{AB}\},}{Eq7}
\noindent where $\Re\{T_{AB}\}$ denotes the real part of $T_{AB}$, and $N_i$ are constants such that $0\leq T_{\varhexagon}\leq 1$. Shifting the origin to a $B$-site atom or honeycomb hollow site amounts to multiplying $T_{AB}$ in Eq. 11 by a phase shift of $e^{-2\pi i/3}$ or $e^{-4\pi i/3}$, respectively. A triangular lattice is recovered when $\alpha$ or $\beta$ is set to zero.

\begin{figure*}[!ht]
	% \centering
	\includegraphics[width=\textwidth]{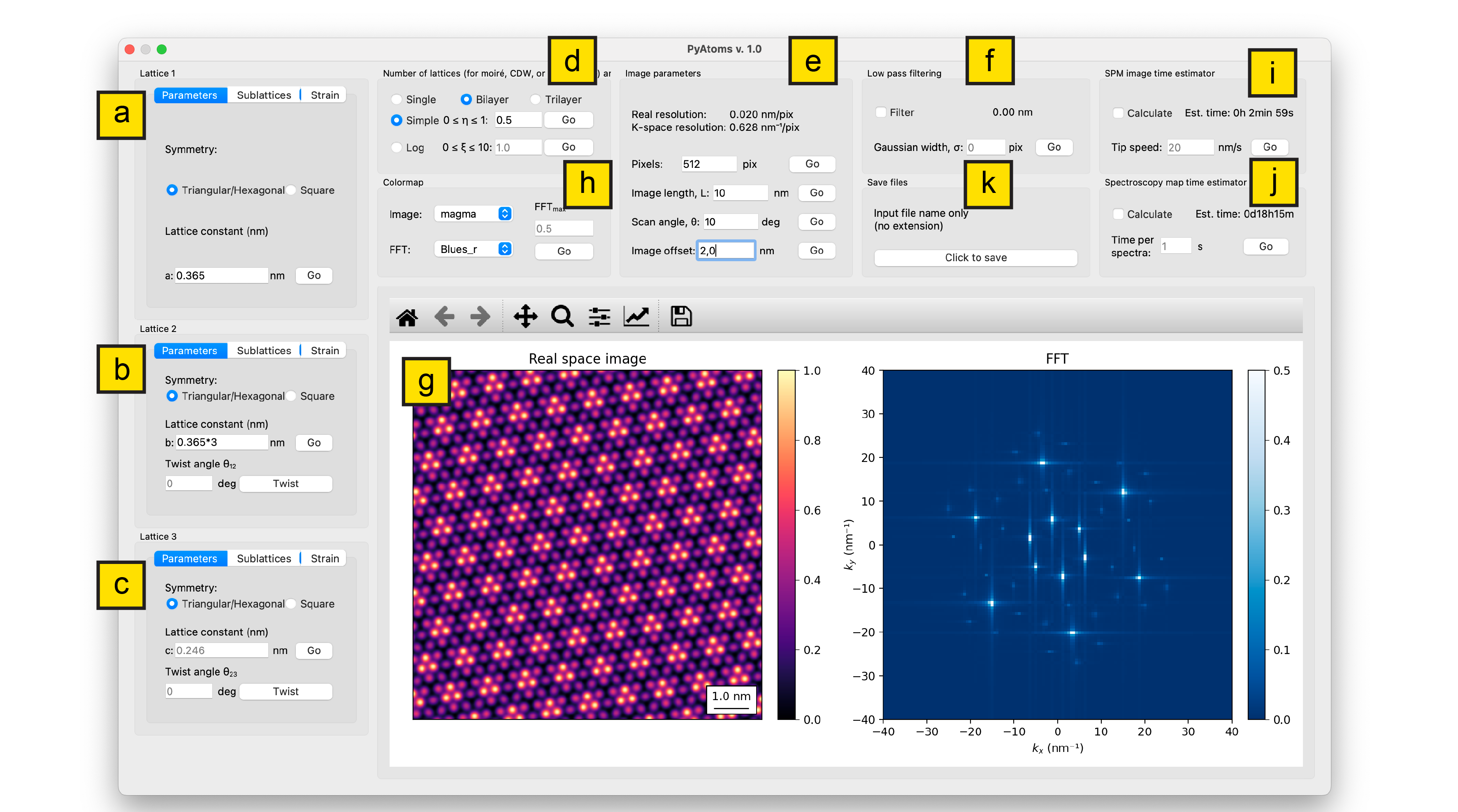}
	\caption{\textbf{PyAtoms graphical user interface.} See text for brief descriptions of each labeled window.} 
	\label{fig:PyAtomsGUI}
\end{figure*}

Figure \ref{fig:sublattice} displays several simulated STM images of a honeycomb crystal centered at a hollow (H) site. Ideal graphene (lattice constant $a=0.246$ nm) is simulated when the amplitudes of the $A$ and $B$ sublattices are equal (Fig. \ref{fig:sublattice}(a)). A vertical line cut of the simulated image (Fig. \ref{fig:sublattice}(b)) shows that the honeycomb crystal (solid black line) is composed of two triangular lattices, with atoms from each lattice centered at $A$ (solid red line) and $B$ (solid blue line). Modifying the relative amplitudes of the two sublattices $(\alpha,\beta)$ can create simulated STM images of more general honeycomb structures (Fig. \ref{fig:sublattice}(c)). This can be used to model STM images of transition metal dichalcogenides, graphene with broken sublattice symmetry, or experimental STM images recorded with an asymmetric tip. The relative sublattice strength $(\alpha, \beta)$ and the image origin ($A$, $B$, or hollow site) is easily implemented using PyAtoms's GUI (Fig. \ref{fig:PyAtomsGUI}(a)). 

As an example of PyAtoms' robust capabilities, consider Fig. \ref{fig:sublattice}(d), which displays an experimental STM image of the surface of graphite. Typically, the surface of graphite appears in STM as a triangular lattice \cite{binnig_scanning_1987,hembacher2003revealing}. The presence of the other sublattice is thus an artifact, most likely due to an asymmetric or doubled STM tip. This is easily simulated with PyAtoms in Fig. \ref{fig:sublattice}(e) by choosing sublattice parameters ($\alpha=0.45,\beta = 0.55$).

\subsection{Superlattices and multi-layered systems}

A superlattice is a larger periodic structure that is superimposed over an existing lattice. Superlattices can emerge from surface reconstructions \cite{binnig_scanning_1987}, alloys and interfaces of materials \cite{carpinelli_direct_1996}, the presence of periodic lattice distortions and/or charge- or bond-density waves \cite{Sacks_PhysRevB.57.13118,arguello_visualizing_2014,pasztor_holographic_2019,Shunsuke_PhysRevLett.132.056401}, or, in moir\'{e} systems, from the combination of two or more crystal lattices \cite{rutter2007scattering,xue2011scanning,decker_local_2011,zhang_interlayer_2017,kerelsky_maximized_2019,jiang_charge_2019,xie_spectroscopic_2019,li_imaging_2021,kim_imaging_2023,slot2023quantum}. The moir\'{e} superlattice (wavevectors $\bvec{g}_{A,i,B,j}$) can be understood as the real-space periodic analog of the beat-frequency in sound: Interference between two periodic lattices (labeled $A$ and $B$ with wavevectors $\bvec{g}_{A,i}$ and $\bvec{g}_{B,j}$, where $i,j$ index the Bragg wavevectors) will create a moir\'{e} superlattice with wavevectors given by the difference, $\bvec{g}_{A,i,B,j} = \bvec{g}_{A,i}-\bvec{g}_{B,j}$. The smaller the difference between the two lattices -- either in magnitude or direction, i.e. twist -- the smaller the moir\'{e} wavevector, and thus the longer the real-space wavelength, $\lambda_M \propto |\bvec{g}_{A,i,B,j}|^{-1}$. The presence of this long-wavelength moir\'{e} periodic potential can have a dramatic effect on the electronic properties of 2D material stacks, from inducing superconductivity to correlated insulating behavior and magnetism \cite{cao_correlated_2018,cao_unconventional_2018,sharpe_emergent_2019,balents2020superconductivity,mak2022semiconductor}.

In PyAtoms, we currently employ two models that simulate the vast majority of scanning probe images of atomic superlattices in both real- and reciprocal-space.

\begin{figure*}[ht]
	\centering
	\includegraphics[width=\linewidth]{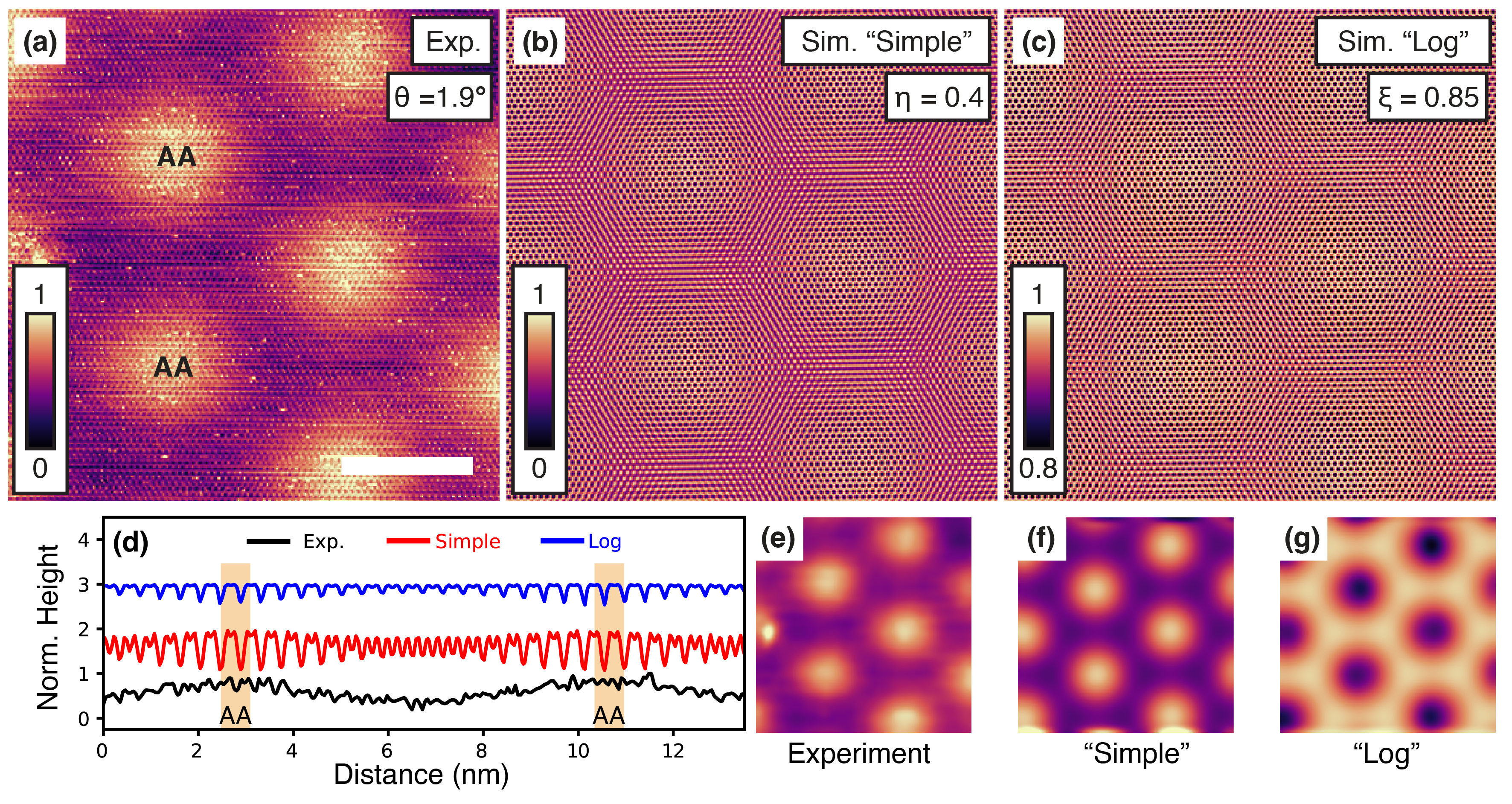}
	\caption{\textbf{Moir\'{e} simulations of experimental data.} (a) Experimental STM image of twisted graphene-graphite with a\moire twist of $\theta=1.9^\circ$. ($V_b = -400$ mV, $I_t = 250$ pA, $T = $ 4.7 K). The height has been normalized to $[0,1]$ . (b) Simulation using the ``simple''\moire model with $\theta_{12}=1.9^\circ,\eta=0.4$. (c) Same as (b) using the ``log''\moire model with $\xi = 0.85$. Note the compressed color scale. (d) Linecuts between the two AA\moire sites in (a)-(c). (e-f)Same as (a-c) after low-pass filtering ($\sigma = 6$ pixels) which highlights the differences between the simulated\moire models. All images have a lateral size of 18.7 nm $\times$ 18.7 nm. 
	}	
	\label{fig:MoireExamples}
\end{figure*}

\noindent\textbf{Toy or ``Simple'' model:} This is a minimal toy model that yields a superlattice that is given by,
\begin{align}
	T_{M}^S \propto (1-\eta)\sum_{l=1} T_l + \eta\prod_{l=1} T_l .
	\label{moire_eqn_toy)}
\end{align}
Here, $T_M^S$ represents the moir\'{e} or other superlattice image, $T_l$ represents the individual constituent lattices, and $\eta$ is a parameter ($0\leq \eta \leq 1$) that weighs the relative strength of the sum of the lattices, $\sum_l T_l$, to their product,  $\prod_l T_l$. Our toy model has notable strengths: (i) a single toy parameter, $\eta$, that is GUI-friendly; (ii) it creates real-space simulations with a wide tuneability in the image contrast to best compare with experimental data; and (iii) low-pass filtering of the resultant image agrees with experimental data (described below and Fig. \ref*{fig:MoireExamples}f). Using this model, the FT-STM image --- by design --- contains only first-order Bragg and superlattice peaks, which is typical of most experimental STM data \cite{gutierrez_imaging_2016,xue2011scanning,decker_local_2011,joucken_fourier_2015}.

\bigskip
\noindent\textbf{Tunneling or ``Log'' model:} An alternative model proposed by Joucken \textit{et al} \cite{joucken_fourier_2015} for simulating multi-layer moir\'{e} superlattices is rooted in the STM constant-current tunneling process and additionally takes into account the inter-atomic layer spacing, $d$, and the out-of-plane exponential decay length, $\lambda$. Here, the STM tunnel current is given by,
\begin{equation}
	I(x,y,z) \propto  e^{-z(x,y)/\lambda}\sum_{l=1}  T_l e^{-(l-1)d)/\lambda},
	\label{tunnel_eqn}
\end{equation}

\noindent where $z(x,y)$ is the STM tip height at lateral position $(x,y)$ and $T_l$ ($l=1,2,...)$ describes the LDOS in the $l$th atomic layer from the surface. In an STM constant-current measurement, the tip height, $z(x,y)$, changes dynamically to maintain a constant tunnel current, $I(x,y,z)$, through a feedback-loop. Setting the left-hand side of Eq. \ref{tunnel_eqn} constant and solving for the tip height, $z(x,y)$, yields the resultant superlattice STM topographic height image, $T^L_M = z(x,y)$, given by 
\begin{equation}
	T^L_M \propto \ln \left|\sum_l T_l e^{-(l-1)d/\lambda}\right|\propto\ln \left|\sum_l T_l e^{-(l-1)\xi}\right|,
	\label{moire_eqn_log}
\end{equation}

\noindent where, in the last step, we have combined the ratio of the inter-layer distance ($d$) to the decay length ($\lambda$) into a single parameter, $\xi = d/\lambda$: $\xi$ small corresponds to a short interlayer distance (or long atomic decay length); $\xi$ large corresponds to a large interlayer distance (or short atomic decay length).  This model has its own notable strengths: (i) a single parameter, $\xi$, that is GUI-friendly; (ii) the relative intensity of constituent lattices can be tuned by adjusting $\xi$, and (iii) the Fourier-transformed image contains all orders of the superlattice peaks. However, the logarithm in Eq. \ref{moire_eqn_log}, strongly affects the simulated image's dynamic range which makes adjusting the contrast to match experimental data much more difficult. By allowing the user to toggle between the two models (``Simple'' or ``Log''), the user can determine the model and parameters to best match their experiment. 

Figure \ref{fig:MoireExamples} contrasts the two models. Shown in Fig. \ref{fig:MoireExamples}(a) is an experimental STM image recorded on graphite at $T=$ 4.7 K. The presence of a\moire pattern indicates the delamination and twisting of the top-most graphene layer \cite{li_scanning_2009,li_observation_2010}. The twist angle can be determined from the\moire wavelength using well-known geometric relations which yields $\theta_{12}\approx 1.9^\circ$. Alternatively, the parameter can be determined using PyAtoms, which can rapidly simulate the data as shown in Fig. \ref{fig:MoireExamples}(b,c) using either the ``Simple'' or ``Log'' model. The key differences between the two models can be seen by taking a linecut connecting the\moire AA sites (Fig. \ref{fig:MoireExamples}(d)). The ``Simple'' model (red line) has an amplitude similar to that in experiment (black line). The ``Log'' model (blue line), however, has a much narrower height range and, owing to Eq. \ref{moire_eqn_log}, additionally features deep minima at the AA sites. Low-pass filtering the experimental and simulated images (Fig. \ref{fig:MoireExamples}(e)-(g)) shows that the ``Simple'' model more closely follows the experimental\moire image while the ``Log'' model yields an inverted filtered image owing to the deep minima at the AA sites.

We note other recent analytical models for simulating \moire and/or multilayered STM images. Le Ster \textit{et al} \cite{le2024moire} have presented the \moire plane wave expansion (MPWE) model which produces a total STM image that includes interactions between layers by weighing the contributions of three terms: (i) a non-interacting substrate layer, $z(\mathbf{r})$ composed of plane waves with wavevectors $\mathbf{G}_{mn}$; (ii) a non-interacting second layer, $z'(\mathbf{r})$ composed of plane waves with wavevectors $\mathbf{G'}_{pq}$; and a \moire term, $z_M (\mathbf{r})$, which takes into account interactions by being composed of plane waves with wavevectors, $\mathbf{K}_{mnpq} = \mathbf{G'}_{pq}-\mathbf{G}_{mn}$, which are all possible linear combinations of wavevectors of the constituent layers \cite{le2019moire}. P\'{a}sztor \textit{et al} \cite{pasztor2024delusive} have presented a similar model which, for the purpose of investigating artifacts in geometric phase analyses, consists of summing (without weights) plane waves with all first-order linear combinations of wavevectors of the constituent lattices. Future versions of PyAtoms can easily allow these and even more models. We stress that, since the goal of PyAtoms is to rapidly simulate \textit{ideal} moir\'{e} and other super-lattices, the simple sums of plane waves cannot account for lattice relaxation that can occur in real systems  \cite{woods_commensurateincommensurate_2014}.

\subsection{Homogeneous strain}

Lattice strain can strongly affect the electronic and optical properties of materials. When a crystal lattice is strained, the atomic bonds in real-space become distorted, consequently shifting the reciprocal lattice vectors. This can distort the shape of the Fermi surface and the Brillouin zone boundaries, affecting the vibrational properties, and even the opening of band gaps in the band structure \cite{pereira_tight-binding_2009}.

To leading order in the strain tensor, $\boldsymbol{\epsilon}$, the presence of strain affects the direct lattice vectors via

\eq{\bvec{v} = (\mathbb{I}+\boldsymbol{\epsilon})\bvec{v}_0,}{Eq9}

\noindent where $\bvec{v}_0$ is the un-strained direct lattice vector, $\bvec{v}$ is the strained direct lattice vector, $\mathbb{I}$ is the $2\times 2$ identity matrix, and $\boldsymbol{\epsilon}$ is the two-dimensional strain tensor given by

\eq{
	\boldsymbol{\epsilon} = 
	\begin{pmatrix}
		\epsilon_{xx} & \epsilon_{xy}\\
		\epsilon_{yx} & \epsilon_{yy}
	\end{pmatrix}	
	.}{Eq8}

In PyAtoms, the strain $\hat{x}$-axis is directed along the first direct lattice vector $\bvec{a}_1=a\hat{x}$. The resulting strain-distorted reciprocal lattice vectors are given by (to leading order in $\boldsymbol{\epsilon}$),
\begin{align}
	\bvec{g}_1^{sq} &\approx \frac{2\pi}{a}[(1-\epsilon_{xx})\hat{x} -\epsilon_{xy}\hat{y}] \\
	\bvec{g}_2^{sq} &\approx \frac{2\pi}{a}[-\epsilon_{xy}\hat{x} + (1-\epsilon_{yy})\hat{y}]
\end{align}
for the square lattice and
\begin{align}
	\bvec{g}_1^{tri} &\approx \frac{2\pi}{a}\Big[\Big(1-\epsilon_{xx}-\frac{\epsilon_{xy}}{\sqrt{3}}\Big)\hat{x} + \Big(\frac{1}{\sqrt{3}} -\epsilon_{xy}-\frac{\epsilon_{yy}}{\sqrt{3}}\Big)\hat{y}\Big] \\
	\bvec{g}_2^{tri} &\approx \frac{2\pi}{a}\Big[\Big(-1+\epsilon_{xx}-\frac{\epsilon_{xy}}{\sqrt{3}}\Big)\hat{x} + \Big(\frac{1}{\sqrt{3}} +\epsilon_{xy}-\frac{\epsilon_{yy}}{\sqrt{3}}\Big)\hat{y}\Big]
\end{align}
for the triangular lattice.

\begin{figure*}[ht]
	% \centering
	\includegraphics[width=\textwidth]{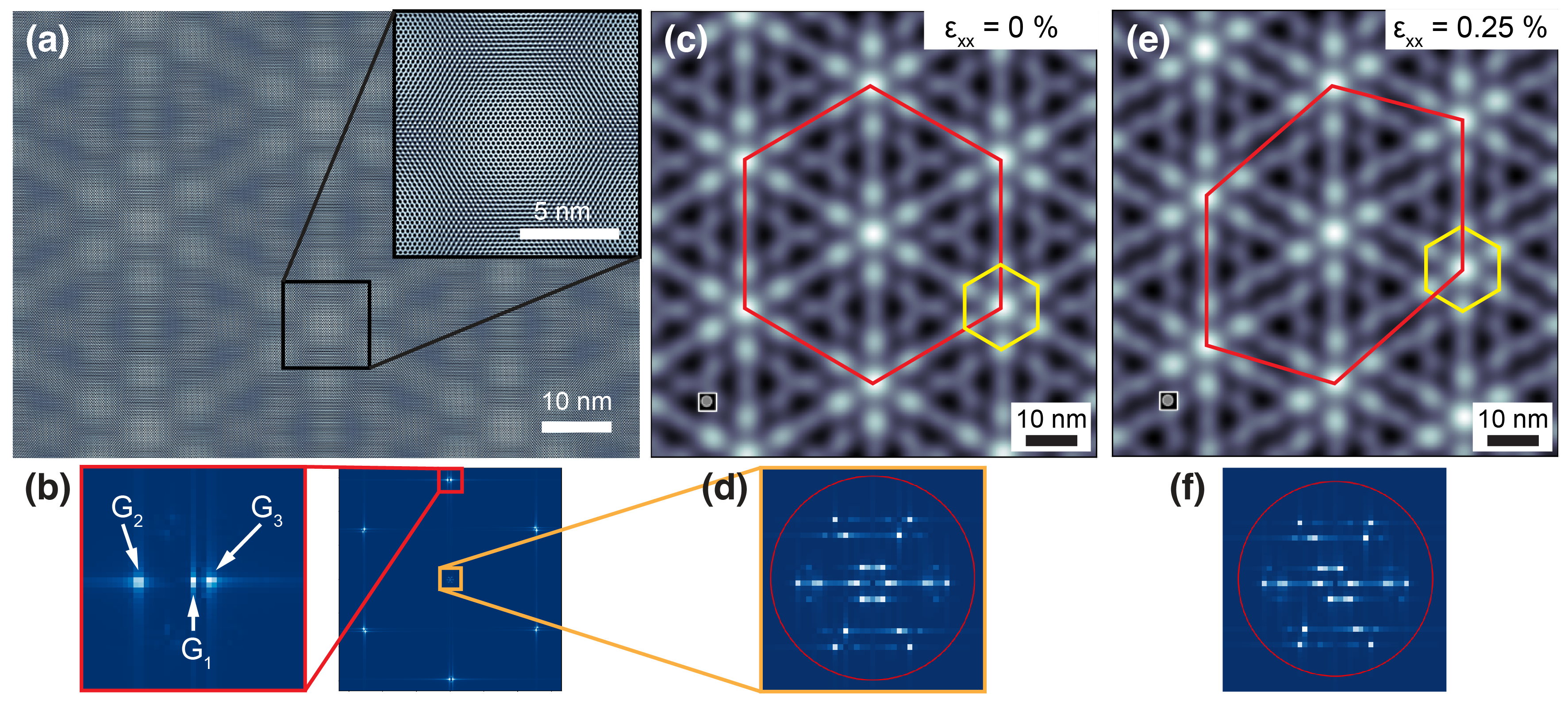}
	\caption{\textbf{Twisted trilayer graphene simulation.} (a) Cropped image of a PyAtoms simulation of unstrained twisted trilayer graphene: 2048 pixels, $a = $ 2.46 \AA, $L = $ 90 nm, $\theta_{12} = 1.42^\circ$, $\theta_{23} = -1.88^\circ$, \moire `Simple' mode. Inset: Atomic-scale zoom on AAA region. (b) (Right) Full-scale FFT of (a) and (Left) detail zoom showing the Bragg peaks for each graphene layer. (c) Low-pass filtered image of (a): $\sigma_R = $ 20 pixels ($0.88$ nm). The solid lines highlight the nearest-neighbors for the long (red) and short (yellow) \moire wavelength. (d) Zoom FFT of (c) near the origin. The circle displays the filtered region in $k$-space. (e) Same as (c) with a small strain on $G_1$ that warps the \moire pattern. (f) FFT of the strained trilayer graphene.} 
	\label{fig:MoireStrain}
\end{figure*}

\subsection{Low-pass filtering}

PyAtoms is equipped with a low-pass Gaussian filter option that eliminates short spatial frequencies in an image while preserving long wavelength periodicities. This option is useful in simulating STM images of moir\'{e} lattices, where it can be used to remove the atomic lattice, which may not be present in experimental data, either due to the scan size or the particular tunneling conditions. The low-pass filter can also be used to remove common aliasing effects present in STM data.

For computational efficiency and speed, the low-pass filter in PyAtoms is performed in reciprocal space by taking the product of a Gaussian mask with reciprocal-space width $\sigma_K$ and the 2D Fourier transform of the atomic image and then inverse Fourier transforming to real-space. The Gaussian mask  is given by
\eq{\Gamma_K = \frac{1}{2\pi\sigma_K}e^{-r^2/2\sigma_K^2},}{Eq_mask} 

% \pagebreak

\noindent where the real-space Gaussian width of the filter is given by $\sigma_R = 1/\sigma_K$.

\begin{figure*}
	% \centering
	\includegraphics[width=\textwidth]{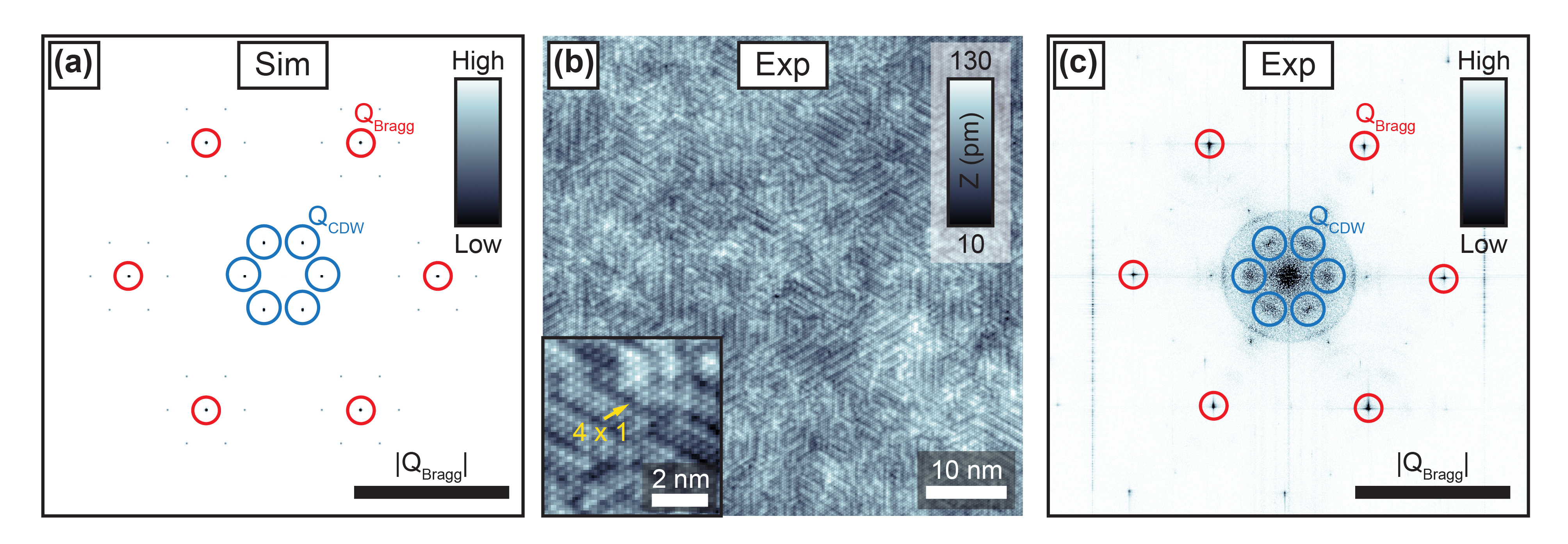}
	\caption{\textbf{Setting up and optimizing QPI images with PyAtoms.} (a) Simulated FT-STM ($624\times 624$ pixels, 60 nm) of NbSe$_2$ with three-fold $(4\times 1)R0^\circ$ CDW domains. (b) Experimental STM topography image of NbSe$_2$ at $T = 4.7$ K with strain-induced $(4\times 1)R0^\circ$ disordered CDW domains ($V_b = 1.5$ mV, $I_t = 3$ pA). Image has optimized scan parameters from (a). Image acquisition time $\approx 90$ min. (c) FT-STM of the topography in (b) showing additional features corresponding to QPI patterns seen previously \cite{gao2018atomic}.}
	\label{fig:NbSe2QPI}
\end{figure*}
    
\section{PyAtoms Overview}

Figure \ref{fig:PyAtomsGUI} displays the PyAtoms interface. Below we briefly describe the functionality of each.

\bigskip
\noindent\textbf{(a, b, c) Lattice 1, 2, 3:} In the \texttt{Lattices} section, users select the properties for each constituent layer. In the \texttt{Parameters} tab users select: the lattice symmetry (triangular/hexagonal or square), which sets the primitive lattice vectors; the lattice constant, in nanometers; and the twist angle ($\theta_{12},\theta_{23}$) for two- or three-layer systems. For triangular/hexagonal systems,  the \texttt{Sublattices} tab allows users to select: the lattice site at the image origin (hollow, A, or B) and the amplitude of each sublattice ($\alpha_i,\beta_i$) (see Fig. \ref{fig:sublattice}). In the \texttt{Strain} tab, users have the option to apply a 2D strain tensor (Eq. 17), where the $x$-axis of strain tensor for $i$th layer is along the direction of the $i$th layer's first reciprocal lattice vector, $\bold{g}_1^i$ (Eqs. 3, 4). 

\bigskip
\noindent\textbf{(d) Number of lattices:} Here, users select the number of layers/lattices to simulate: single, bi-, or tri-layer lattice as well as the model for approximating the combined system (``Simple'' or ``Log'') and the corresponding parameters ($\eta,\xi$).

\bigskip
\noindent\textbf{(e) Image parameters}: This section sets the configuration and resolution of the entire image and closely follows typical parameters in STM data collection software. Key image parameters include the number of pixels, the length of the image (in nanometers) and the scan angle, $\theta$, which is along the first reciprocal lattice vector of the 1st layer, $\bold{g}^1_1$, and the image offset, $(x_0,y_0)$, where both are real numbers that set the origin of the image. The image parameters tab features a real-time calculation of the resolution in real-space, defined by $\delta R = L/(N_{pix}-1)$, where $N_{pix}$ is the number of pixels, and the reciprocal space resolution given by $\delta k = 2\pi/L$. These values are useful when using PyAtoms for preparing for time-consuming QPI measurements in order to make sure that the pixel sampling rate is above the Nyquist frequency.

\begin{figure}[ht]
	\centering
	\includegraphics[width=\linewidth]{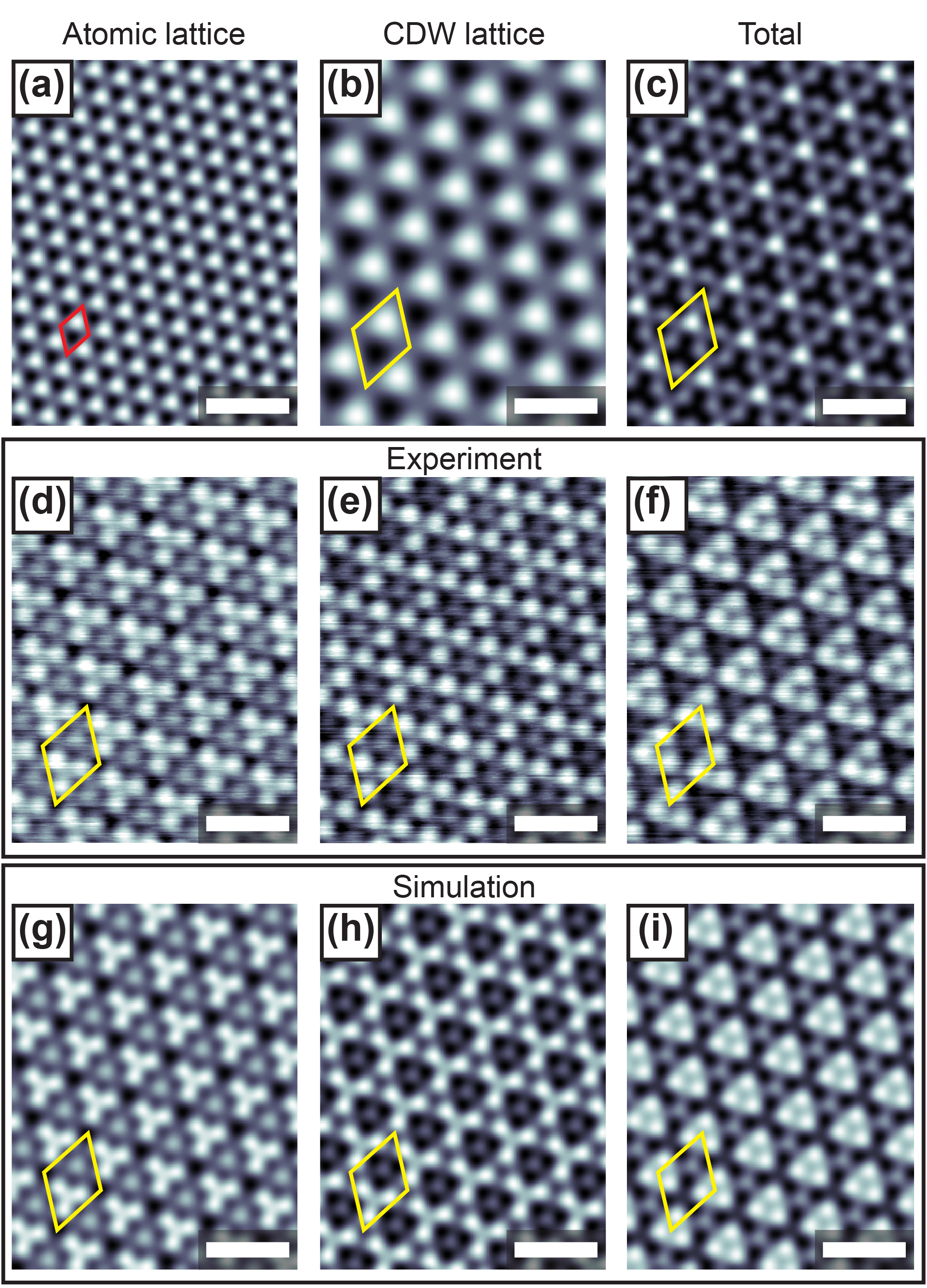}
	\caption{\textbf{Simulating bias-dependent $\boldsymbol{(2\times 2)R0^\circ}$ CDWs in a transition metal dichalgogenide.} (a) PyAtoms simulation of NbSe$_2$ atomic lattice using sublattice parameters $\alpha_{at}=0.7,\beta_{at}=0.3$ and (b) the $(2\times 2)R0^\circ$ CDW lattice using sublattice parameters $\alpha_{cdw}=0.7,\beta_{cdw}=0.3$. (c) Simulation of the combined atomic and CDW lattices using the `Simple' option ($\eta=0.2$). (d)-(f) Experimental STM topographs of uniaxial strain-induced commensurate $(2\times 2)R0^\circ$ CDW (diamond) in NbSe$_{2}$ recorded at $T = 4.7$ K and $I_t$ = 200 pA. Voltage biases, $V_b$, are (d) $-1.3$ V, (e)$ -1.4$ V, and (f) $-1.6$ V. Corresponding PyAtoms simulations are shown in (g)-(i) using the `Simple' option and CDW sublattice parameters (g) $\alpha_{cdw}=0.77,\beta_{cdw}=0.23$, (h) $\alpha_{cdw}=0.55,\beta_{cdw}=0.45$, (i) $\alpha_{cdw}=0.98,\beta_{cdw}=0.02$. All scale bars are 1 nm.} 
	\label{fig:strain_NbSe2}
\end{figure}

\bigskip
\noindent\textbf{(f) Low pass filtering}: For multilayer simulations, users have the option to input a value for \texttt{sigma}, the radius of Gaussian mask in real space (pixels) units. The full-width at half-max (FWHM) of the Gaussian filter is shown as a white circle in the bottom left corner of the real space image. Likewise, the FWHM of the Gaussian filter in reciprocal space is shown as a red circle in the FFT (Fig. \ref{fig:MoireStrain}(d,f)). The radius of the Gaussian mask in nanometers is shown in real-time in the text box.

\bigskip
\noindent\textbf{(g) Imaging window}: This window displays the real-time simulated real-space and reciprocal-space images of the resultant lattice system. The navigation buttons allow the user to zoom, pan, and tune each plot's image scale. The save button outputs a screenshot of this imaging window, including all axes and text.

\bigskip
\noindent\textbf{(h) Colormap}: This section enables users to choose the colormap for the real space and FFT images. Additionally, users have the option to set the maximum value of the FFT colormap to enhance less pronounced features.

\bigskip
\noindent\textbf{(i) SPM image time estimator}: This section estimates the time required to complete scanning of an image, using the values in the \texttt{Parameters} tab along with the scan/tip speed, $v_t$ (in nanometers/second).  This is calculated assuming the SPM tip scans in the left/right fast-scan direction and in a single slow-scan direction (upwards, for example). This time is estimated via \eq{T_{im} = {2N_{pix}L}/{v_t}}, where $N_{pix}$ is the number of pixels in one line, $L$ is the length of the image in nanometers, $v_t$ is the velocity (in nm/s) of the tip in one direction and the factor of 2 takes into account both the left and right fast-scan directions. This section is useful in estimating the time to record a ``closed-feedback'' $dI_t/dV_b$ map where the LDOS($eV_b$) is recorded at a single energy as the tip slowly scans the surface.

\bigskip
\noindent\textbf{(j) Spectroscopy map time estimator}: This section estimates the time required to complete a pixel-by-pixel square grid spectroscopy map using the values in the \texttt{Parameters} tab along with the time (in seconds) that each spectra requires, $T_{spec}$. We also assume that spectra are recorded in one direction along both the fast- and slow-scan direction. The time is estimated via
\eq{T_{map} = N^2_{pix}T_{spec} + 2N_{pix}L/{v_t}}. The user inputs the time per spectra, which is the total time in seconds (including system overhead) to record a single spectrum, $T_{spec}$, i.e., one $dI_t/dV_b$ sweep. This section is useful in estimating the time to record an ``open-feedback'' LDOS($eV_b$) map where a $dI_t/dV_b(V_b)$ spectrum is recorded at every pixel position. Together with the imaging window (Fig. \ref{fig:PyAtomsGUI}(g)) provided by PyAtoms, this estimator is invaluable in ensuring that time-consuming QPI measurements will have the requisite reciprocal space resolution and spatial extent while avoiding aliasing effects.

\bigskip
\noindent\textbf{(k) Save Files}: This operation generates a folder named \texttt{FileName} containing four files: a PNG file of the real space image, a PNG file of the FFT, a text file containing the key PyAtoms parameters, and a tab-delimited text file containing the real space image in a 2D array that can easily be imported for further analysis.

\section{Applications}

Owing to its simple and real-time GUI interface, PyAtoms is a useful research tool both during STM data collection as well as for data analysis. As shown in Fig. \ref{fig:MoireExamples}, PyAtoms can be used side-by-side during STM measurements in order to rapidly simulate experimental images and determine properties such as the constituent lattice constants and\moire twist angles between layers. Below we describe some examples of PyAtoms research capabilities.

In practice, the strain tuning capabilities of PyAtoms can be used to simulate and determine either the amount of true, physical strain in STM images or the apparent strain due to scan artifacts such as tip drift. Figure \ref{fig:MoireStrain} demonstrates the dramatic effect that strain can have on the\moire lattice and how PyAtoms can be used to determine this strain. Fig. \ref{fig:MoireStrain}(a) is a simulation of twisted trilayer graphene ($\theta_{12}=1.42^\circ, \theta_{23}=-1.88^\circ$)\moire quasicrystal \cite{uri_superconductivity_2023} that features two\moire lattices. The atomic Bragg peaks for each graphene layer can be seen in the Fourier transform (Fig. \ref{fig:MoireStrain}(b)). The complex spatial structure of the\moire-of-\moire pattern (Fig. \ref{fig:MoireStrain}(c)) can be better seen after low-pass filtering to keep only the long wavelength periodicities (Fig. \ref{fig:MoireStrain}(d)). The presence of even miniscule strain in the top-most graphene layer ($\epsilon_{xx}=0.25$\%) significantly alters the larger\moire pattern in both real (Fig. \ref{fig:MoireStrain}(e)) and reciprocal space (\ref{fig:MoireStrain}(f)). For scanning probe researchers studying\moire multi-layers, PyAtoms can be used side-by-side during measurements to quickly determine the local strain in an area so that users can make informed decisions whether to seek other, less-strained areas.

Next, we show how PyAtoms can be used to determine and optimize scan parameters for performing QPI measurements, which reveal subtle scattering patterns at wavevectors  that connect different portions of constant energy contours of the electronic band structure \cite{hoffman_imaging_2002,avraham_quasiparticle_2018}. Here, we focus on NbSe$_2$ which undergoes a strain-induced transition from an incommensurate $(3+\delta)\times(3+\delta)R0^\circ$ CDW to local areas of commensurate $(4\times 1)R0^\circ$ CDW order \cite{gao2018atomic}. PyAtoms' GUI allows users to view and optimize desired scan parameters to achieve high $k$-space resolution ($\delta k = 2\pi/L$, image length $L$ large), sufficient $k$-space extent ($k_{max} = \frac{1}{2}N\delta k = \pi N/L = 1.25 |Q_{\text{Bragg}}|$, and image acquisition time, $T\approx 90$ min (Eq. 23), as shown in Fig. \ref{fig:NbSe2QPI}(a). Using these optimized parameters, the experimental real space image can be recorded at low tunneling bias (Fig. \ref{fig:NbSe2QPI}(b)) which, through its FFT (Fig. \ref{fig:NbSe2QPI}(c)) provides a QPI image near the Fermi energy \cite{joucken2020determination}. 

PyAtoms is additionally useful in simulating real-space CDW systems of various geometries and phases. Shown in Fig. \ref{fig:strain_NbSe2} are experimental low-temperature STM topographic images of 2$H$-NbSe$_2$ with uniaxial strain applied along the vertical direction which has created a local transition from an incommensurate $(3+\delta)\times(3+\delta)R0^\circ$ CDW \cite{moncton_neutron_1977,soumyanarayanan2013quantum,arguello_visualizing_2014,gye2019topological} to commensurate $(2\times 2)R0^\circ$ CDW order, in agreement with previous strain-STM experiments\cite{gao2018atomic}. The phase of the CDW in the STM images in Fig. \ref{fig:strain_NbSe2}(a)-(c) is strongly energy-dependent and strikingly similar to STM topographs of $1T$-TiSe$_2$ \cite{pasztor_holographic_2019}.   

\begin{figure*}
	\centering
	\includegraphics[width=\linewidth]{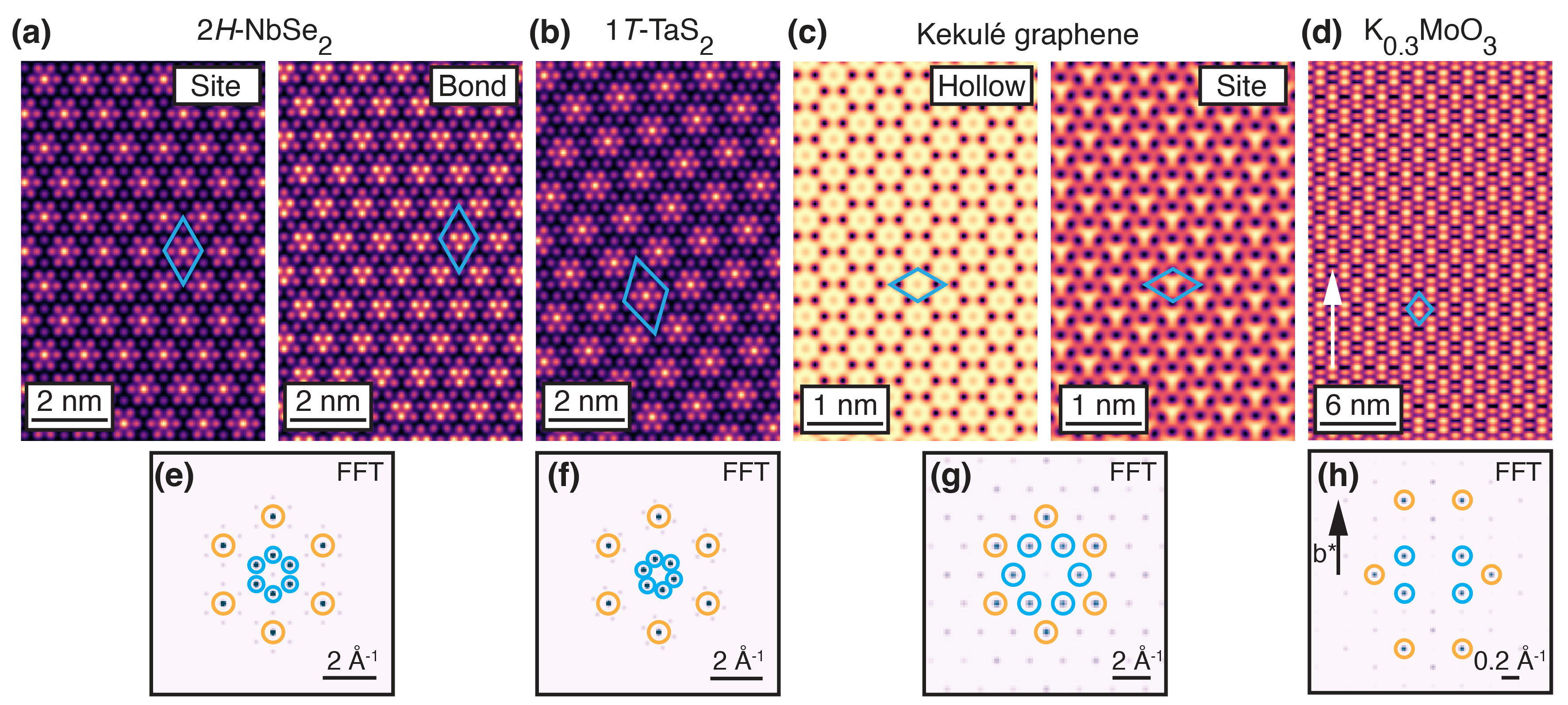}
	\caption{\textbf{Site- and bond-centered charge density wave simulations.} (a) Simulation of the approximately $(3\times 3)R0^\circ$ CDW in 2$H$-NbSe$_2$ ($a=3.44$ \AA) which can be Se site- (left) or bond-centered (right). (b) ``Star-of-David'' $(\sqrt{13}\times\sqrt{13})R13.9^\circ$ CDW in 1$T$-TaS$_2$ ($a = 3.36$ \AA). (c) Kekul\'{e}-distorted graphene with $(\sqrt{3}\times\sqrt{3})R30^\circ$ bond-density wave texture which can be hollow- (left) or C site-centered (right).  (d) Incommensurate CDW in quasi-1D  K$_{0.3}$MoO$_3$, potassium blue bronze. (e)-(h) The respective Fourier transforms. The atomic peaks (yellow) and CDW peaks (blue) are highlighted. (a)-(b) simulated with ``Simple'' model ($\eta = 0.3$); (c) and (d) simulated with ``Log'' model using $\xi = 0$ and $\xi = 1$, respectively.}	
	\label{fig:SimulationExamples}
\end{figure*}

PyAtoms can model a host of CDW orders in transition metal dichalcogenide (TMDC) van der Waals systems \cite{wilson_charge-density_1975}. Figure \ref{fig:SimulationExamples}(a,e) displays a simulation of 2$H$-NbSe$_2$, a TMDC that hosts a nearly commensurate CDW \cite{moncton_neutron_1977} with order that is approximately $(3\times 3)R0^\circ$ and is shown with a site-centered (left) and bond-centered (right) CDW phase. Figure \ref{fig:SimulationExamples}(b,f) displays the $(\sqrt{13}\times\sqrt{13})R13.9^\circ$ CDW of TMDC $1T$-TaS$_2$. Figure \ref{fig:SimulationExamples}(c,g) displays simulations of graphene with a $(\sqrt{3}\times\sqrt{3})R30^\circ$ BDW, termed a Kekul\'{e} distortion with a phase centered at a hollow site (``Kekule-O,'' left) \cite{li_scanning_2019,liu_visualizing_2022} or carbon site (``Kekul\'{e}-Y,'' right) \cite{gutierrez_imaging_2016,nuckolls_quantum_2023}. Finally, we show that PyAtoms can also be used to model the CDW in quasi-1D Peierls system K$_{0.3}$MoO$_3$, potassium blue bronze \cite{travaglini_blue_1981,brun_charge-density_2005}. Fig. \ref{fig:SimulationExamples}(d) shows the monoclinic atomic lattice projected onto the $(\bar{2}01)$ surface, which is what is measured by STM \cite{brun_charge-density_2005}. The surface structure is simulated by a strained triangular lattice ($a_1 = 7.67$ \AA, $\epsilon_{yy}=32\%$, image scan angle $\theta = 90^\circ$). The nearly commensurate CDW, has a projection along the vertical chain direction (white arrow) ($|\bvec{q}^b_{CDW}|/|\bvec{b}^*|\approx 0.25$) for the ($\bar{2}01$) surface \cite{travaglini_blue_1981,brun_charge-density_2005} and is simulated with a strained square lattice ($b_1 = 1.39$ nm, $\epsilon_{xx}=\epsilon_{xy}=\epsilon_{yy}=18$\%, $\theta_{12}=45^\circ$). This robust capability to simulate several families of CDW/BDW systems is especially useful when combined with the spectroscopic map time estimators (Fig. \ref{fig:PyAtomsGUI}(i,j)) which allows users to efficiently plan detailed spectroscopic QPI measurements of quantum materials.
   
\section{Conclusion}

We have presented PyAtoms, an interactive open-source Python-based software for the real-time simulation of STM images of 2D materials. With its quick, intuitive, and robust simulation capabilities, we envision PyAtoms as a valuable companion during both STM measurements and STM data analysis. Future PyAtoms developments include offering additional methods for approximating images of multilayer systems\cite{le2024moire}, allowing generic one- and two-dimensional lattices, and, for the CDW/BDW simulations, full control of the phase in real space which would allow for improved simulation of measured STM data \cite{arguello_visualizing_2014,pasztor_holographic_2019,Shunsuke_PhysRevLett.132.056401}. A current weakness in PyAtoms is that the comparison to experiment is largely qualitative. However, apart from its GUI, the open-source code that underlies PyAtoms can be called separately (\textit{e.g. }through a Python terminal) and can be used, for example, to simulate\moire and superlattice systems beyond three layers. In principle, this ability to call on PyAtoms' base functions offers the possibility to determine the best simulation parameters (lattice constant, A- and B-sublattice amplitude, strain) through a least-squares fitting to experimental STM data. This latter capability would elevate PyAtoms to a quantitative tool in the study of 2D materials. As an open-source software that allows for easy user-driven updates, we foresee PyAtoms becoming a valuable tool to the quantum materials scanning probe community.

\begin{acknowledgments}
The authors thank F. Joucken, K. Palot\'{a}s for useful discussions and J. Berger, C. Irving for technical assistance with STM assembly. A.G.P acknowledges financial support from the National Science Foundation Graduate Research Fellowship Program (NSF-GRFP), the UCLA office of the Dean of Physical Sciences through the UCLA Physics Bridge Program, and the UCLA Graduate Division through a Eugene Cota-Robles Fellowship. K.-Y.W. acknowledges financial support from the Julian Schwinger Foundation. C.G. acknowledges teaching relief support from the UCLA offices of the Executive Vice Chancellor and Provost, Vice Chancellor for Research and Creative Activities, and the Vice Chancellor for Academic Personnel and partial support from the UCLA Hellman Society of Fellows. STM measurements were supported by UCLA Division of Physical Sciences startup funds. Uniaxial strain measurements were supported by a UBC-UCLA Collaborative Research Mobility Award via the Office of the Vice Chancellor for Research and Creative Activities at UCLA.  
\end{acknowledgments}
\section*{Author declarations}
\subsection*{Conflict of Interest}
The authors have no conflicts to disclose.
\subsection*{Author contributions}
C.G. created the original PyAtoms base codes (\texttt{hexatoms, squareatoms, moirelattice}) in Matlab. A.G.P. translated the base codes from Matlab to Python and created the GUI interface with input from C.G. Both A.G.P. and C.G. contributed to PyAtoms' later updates. A.G.P., M.I.M.S., K.-Y.W., S.N., A.N., A.E.-I., and C.G. performed STM measurements. M.I.M.S., A.E.-I., and S.B. designed, assembled, and tested the STM uniaxial strain device. C.G. wrote the manuscript and created the figures with input from A.G.P and M.I.M.S. All authors contributed to editing the manuscript.

\section*{Data Availability Statement}
The data that support the findings of this study are available from the corresponding author upon reasonable request.

\bibliography{PyAtoms_UTF8_v6}% Produces the bibliography via BibTeX.

\end{document}